\documentclass[12pt]{article}
\usepackage{amsmath}
\usepackage{amssymb}
\usepackage{amscd}
\usepackage{graphics}
\usepackage{graphicx}
\usepackage{epsfig}      
\usepackage{color}            
\usepackage{epic}             
\usepackage{eepic}            
\usepackage{rotating}         
\usepackage{type1cm}  
\begin{document}

\renewcommand{\thefootnote}{\fnsymbol{footnote}}
\setcounter{footnote}{2}

\title{Backreacted T-folds \\ and \\ Non-geometric regions  in
configuration space}

\author{Waldemar Schulgin$^{1,2}$ and Jan Troost$^{1}$ }

\date{}
\maketitle

\begin{center}
 \emph{$^{1}$Laboratoire de Physique Th\'eorique}\footnote{Unit\'e Mixte du CNRS et
    de l'\'Ecole Normale Sup\'erieure associ\'ee \`a l'universit\'e Pierre et
    Marie Curie Paris 6, UMR
    8549. LPTENS-08/49},
\emph{ \'Ecole Normale Sup\'erieure,  \\
24 rue Lhomond, 75231 Paris Cedex 05, France}\\
{\texttt{Waldemar.Schulgin,\ Jan.Troost @ lpt.ens.fr}}
\vspace{5mm}

\emph{$^{2}$Laboratoire de Physique Th\'eorique et Hautes
Energies, \\CNRS UMR 7589,
Universit\'e Pierre et Marie Curie Paris 6, \\4 place Jussieu, 75252 Paris cedex 05, France}\\
\end{center}

\begin{abstract}
We provide the backreaction of the T-fold doubly T-dual to a
background with NSNS three-form flux on a three-torus. We extend the
backreacted T-fold to include cases with a flux localized in one out
of three directions. We analyze the resulting monodromy domain walls
and vortices. In these backgrounds, we give an analysis of the action
of T-duality on observables like charges and Wilson surfaces. We analyze arguments
for the existence of regions in the configuration space of second
quantized string theory that cannot be reduced to geometry. Finally, by allowing for 
space-dependent moduli, we find a supergravity solution which is a T-fold with hyperbolic monodromies.
\end{abstract}

\renewcommand{\thefootnote}{\arabic{footnote}}
\setcounter{footnote}{0}
\section{Introduction}
In this paper, we study T-folds and their T-dual backgrounds
\cite{Scherk:1979zr}-\cite{Wecht:2007wu}.
T-folds provide generalizations of manifolds. They consist of patches
that can be glued not only by diffeomorphisms, but also by other
symmetries of string theory, in particular elements of the T-duality
(or of the U-duality) group. These generalizations of manifolds may
allow us to considerably enlarge the set of vacua in string theory. In
particular they may find applications in string theory
cosmology (see e.g. \cite{Kounnas:2007pg}) and string phenomenology.
See e.g. the references  \cite{Bouwknegt:2003vb}-\cite{Grana:2008yw} for
 interesting studies of the topology and geometric structure of T-folds,
as well as their behaviour under T-duality.

In the present paper, we firstly
 wish to study a simple class of T-folds in which we control fully the
backreacted geometry. Our first class of T-folds will be T-dual to known
supergravity solutions, which will allow us to determine the fully backreacted
T-fold. The geometry will lay bare further interesting
properties of T-folds as well as some subtleties
associated to their existence.

Moreover, we study how the T-duality map acts on various observables in the
theory, from an original geometric background to its
twisted torus T-dual as well as to the doubly T-dual T-fold. These observables
will include charge, Wilson surfaces, monodromies and curvature.

We analyze in more detail when  T-folds
cannot be put into geometric form under any T-duality
transformation. That is important, since otherwise, after dividing out
the gauge group in second quantized string theory, it would suffice to
integrate over geometric backgrounds.

We then continue to analyze solutions of string theory that are T-folds, and allow for moduli
varying in space. In that way we can construct a new non-trivial example which solves the supergravity
equations of motion and which is a T-fold with hyperbolic monodromies.

\section{The supergravity backreaction}
\label{general}
One way to construct a T-fold is to start out with a space-time
which is a manifold with a three-torus factor $T^3$ and with constant
NSNS three-form flux $H_{(3)}$ on the three-torus. To obtain a T-fold
one applies T-duality along two isometry directions of the
three-torus \cite{Kachru:2002sk}. One exchanges a geometric background for a non-geometric
one. While this T-fold does not extend the space of inequivalent
string theory vacua, the construction is useful to get to grips with
the non-geometry of T-folds, and the associated observables. The hope
is that the lessons we learn can be applied to T-folds (or U-folds)
with no geometric equivalent.  We will study this well-known example,
include its backreaction in our study, comment on its microscopic
origin, provide new observables that are non-trivial after
backreaction and study a subtlety associated to Wilson surfaces.

In this section, we concentrate on the backreaction in this
T-fold background as well as some closely related ones, in a
geometric, twisted torus and T-fold duality frame.

\subsection*{The supergravity equations}
We want to embed a three-torus factor with constant NSNS three-form $H_{(3)}$
into a full string theory background, and extend the example to other
backgrounds with purely NSNS flux. Since the three-form field strength
provides for a non-trivial energy density on the three-torus, we will need to
take into account its backreaction in order to satisfy the equations of motion
of string theory which reduce to the supergravity equations at first order in
the string coupling, and at weak curvature. Since we have a non-trivial
magnetic NSNS three-form flux, the solution carries NS5-brane charge, and we
will therefore take a minimal approach of constructing it using smeared
NS5-branes \cite{Callan:1991at} only. 
There are alternative embeddings that turn on RR-fluxes
\cite{Marchesano:2007vw}.
 
It is known (see e.g. \cite{Cowdall:1998bu})
 that the following background solves
the supergravity equations of motion (universally for type II, type I
and heterotic supergravities):
\begin{eqnarray}\label{original}
ds^2&=&ds^2_{{\mathbb R}^{5,1}}+h\left(\left(dx^6\right)^2+\left(dx^7\right)^2
+\left(dx^8\right)^2+\left(dx^9\right)^2\right) \ ,\nonumber\\
e^{2\phi}&=&h e^{2\phi_0}\ , \nonumber\\
H_{(3)}&=&*dh \ ,
\end{eqnarray}
where the function $h$ is harmonic
 and the hodge star operator acts on
the four-dimensional transverse space parameterized by the coordinates
$x^{6,7,8,9}$.  More precisely the function $h$ is harmonic up to a
source term which is provided by the positions of NS5-branes that
stretch along the six-dimensional space $\mathbb{R}^{5,1}$.  To generate a
three-torus, we compactify the directions $x^7,x^8,x^9$. The direction
parameterized by $x^6$ is the only non-compact direction orthogonal to
the NS5-branes. Thus the function $h$ will be a harmonic function on
$\mathbb{R} \times T^3$.

Our plan is to perform two T-duality transformations along two
isometry directions of the original background to generate a
T-fold \cite{Kachru:2002sk}. To generate isometries, we study  configurations of
NS5-branes that are smeared along a two-torus $T^2$ inside the
transverse space $\mathbb{R} \times T^3$. We choose the directions of the
two-torus to be parameterized by $x^8$ and $x^9$. The harmonic
function will be constant along these directions. It can depend
on the coordinates $x^6,x^7$. We thus extend the set of examples to include
cases with varying flux.

In the course of the next sections, we will study various configurations with
the above properties and it will be convenient to treat them all at
once. Below we study the supergravity equations of motion in such
backgrounds, including their source terms, since it will provide us
with a handle on what happens to the sources after
T-duality. That will give an indication of the microscopic origin of
T-folds.

The supergravity equations of motion become:
\begin{eqnarray}
R_{AA}-\frac{1}{4}H_{A\rho\sigma}{H_A}^{\rho\sigma}+2\nabla_A\nabla_A\phi&=&-\frac{\Delta
  h}{2h} 
\ \ \ {\rm for}\ A=x^6,x^7,x^8,x^9 \ , \nonumber\\
R_{\mu\nu}-\frac{1}{4}H_{\mu\rho\sigma}{H_\nu}^{\rho\sigma}+2\nabla_\mu\nabla_\nu\phi&=&0  
\ \ \ \ {\rm otherwise}  \ ,\nonumber\\
4(\nabla \phi)^2-4\Box\phi-R+\frac{1}{12}H^2&=&\frac{\Delta h}{h^2} \ ,\nonumber\\
dH_{(3)}=d*dh&=&\Delta h\ dx^6 \wedge dx^7 \wedge dx^8 \wedge dx^9\ .
\end{eqnarray}
Let's recall the sources we should associate to the original geometric background.
A source term proportional to the transverse Laplacian $\Delta$ of the function $h$
appears at the position of the NS5-branes. It codes the mass of the
NS5-branes as well as their magnetic charge under the NSNS three-form
flux.  One concrete way to measure the geometric backreaction on the
space due to the presence of the massive NS5-branes is through the
non-trivial scalar curvature (which is a gauge invariant observable on manifolds):
\begin{equation}\label{scalarcurvature}
R=\frac{3}{2h^3}\left((\partial_{6}h)^2+(\partial_{7}h)^2\right)-\frac{3\Delta h}{h^2} \ .
\end{equation}
We turn to the T-dual backgrounds.
\subsection*{The T-dual twisted torus}
To analyze the microscopic origin of the backreacted twisted torus we
compute the source term after one T-duality transformation.  After
performing a T-duality transformation \cite{Buscher:1987qj}\cite{Giveon:1994fu} 
along the direction parameterized
by the coordinate $x^8$, we obtain a background where the embedded
$T^3$ has the topology of a twisted torus \cite{Hull:1998vy,Kachru:2002sk}:
\begin{eqnarray}
ds^2&=&ds^2_{{\mathbb R}^{5,1}}+h\left(\left(dx^6\right)^2+\left(dx^7\right)^2+\frac{1}{h^2}\left(dx^8-bdx^9\right)^2+\left(dx^9\right)^2\right) \ ,\nonumber\\
e^{2\phi}&=&e^{2\phi_0} \ ,\nonumber\\
B_{(2)}&=&0 \ .
\end{eqnarray}
The value of the NSNS two-form potential $B_{(2)}$ along the isometry
directions in the original background is denoted by $b$, and we have chosen
the other components to be zero\footnote{We will come back to this choice of Wilson surfaces
later.}.
The T-duality transformation exchanges the complex structure modulus
$\tau
$ of the two-torus in the $x^8,x^9$ directions $T^2_{89}$ with its K\"ahler
modulus $\rho=\int_{T^2_{89}}B_{89}+i V_{T^2_{89}}$. 
 Since
we chose the torus to be rectangular,
the dual $B$-field is zero.  Since
in the dual background the NSNS three-form flux $H_{(3)}$ field
vanishes and the dilaton is constant, the supergravity equations of motion 
in the twisted torus background reduce
to equations for the Ricci curvature:
\begin{eqnarray}
&&R_{AA}=-\frac{\Delta h}{h} \ \ \ {\rm for}\ A=x^6,x^7 \ ,\nonumber\\
&&R_{88}=-\frac{\Delta h}{2h^3} \ , \ \ R_{99}=-\frac{(b^2-h^2)\Delta h}{2h^3} \ , \nonumber\\
&&R_{89}=-\frac{b\Delta h}{2h^3} \ , \ \ R_{\mu\nu}=0 \ \ \ {\rm otherwise}  \ .
\end{eqnarray}
and
\begin{equation}
R=-\frac{\Delta h}{h^2}.
\end{equation}
Again we can identify the source terms, which are now purely geometric
singularities. We will discuss them further later on on a case-by-case basis.

\subsection*{The doubly T-dual T-fold}
To generate a backreacted T-fold, we perform a second T-duality transformation
along the $x^9$-direction and obtain expressions for the metric, dilaton and $B$-field:
\begin{eqnarray}
ds^2&=&ds^2_{{\mathbb R}^{5,1}}+h\left(\left(dx^6\right)^2
+\left(dx^7\right)^2+\frac{1}{b^2+h^2}
\left(\left(dx^8\right)^2+\left(dx^9\right)^2\right)\right) \ ,\nonumber\\
e^{2\phi}&=&\frac{h e^{2\phi_0}}{b^2+h^2}\ , \nonumber\\
B_{(2)} &=&-\frac{b}{b^2+h^2} dx^8 \wedge dx^9 \ .
\end{eqnarray}
The local equations of motion in the T-fold background become 
\begin{eqnarray}\label{eomT}
&&R_{AA}-\frac{1}{4}H_{A\rho\sigma}{H_A}^{\rho\sigma}+2\nabla_A\nabla_A\phi
=-\frac{\Delta
  h}{2h} 
\ \ \ {\rm for}\ A=x^6,x^7\ ,\nonumber\\
&&R_{AA}-\frac{1}{4}H_{A\rho\sigma}{H_A}^{\rho\sigma}+2\nabla_A\nabla_A\phi
=-\frac{b^2-h^2}{\left(b^2+h^2\right)^2}
\frac{\Delta h}{2h} \ \ \ {\rm for}\ A=x^8,x^9\ ,\nonumber\\
&&R_{\mu\nu}-\frac{1}{4}H_{\mu\rho\sigma}{H_\nu}^{\rho\sigma}+2\nabla_\mu\nabla_\nu\phi
=0 \ \ \ {\rm otherwise} 
\end{eqnarray}
and
\begin{equation}
4(\nabla \phi)^2-4\Box\phi-R+\frac{1}{12}H^2=\frac{\Delta h}{h^2}\ .
\end{equation}
The scalar curvature associated to the metric is: 
\begin{equation}
R=\frac{3}{2h^3}\left((\partial_{6}h)^2+(\partial_{7}h)^2\right)
+\frac{h^2-3b^2}{\left(b^2+h^2\right)}\frac{\Delta h}{h^2} \ .
\end{equation}
In the following section, we apply the above set of formulas that specify the
backgrounds T-dual to purely NSNS backgrounds. We recall that 
in the geometric setting, we have
parallel NS5-branes distributed evenly over a two-torus at least.


\section{The backreaction and observables: examples}
We turn to concrete examples of NS5-brane backgrounds and their T-duals
to which we apply the above formalism. The examples we study include the
original example of the constant magnetic three-form flux. We
generalize it to include non-trivial values for B-field Wilson lines
(or Wilson surfaces), and we extend it to an example in which we have
a magnetic flux that is uniform only in two directions, and localized
in a third. We examine the domain of validity, the observables and the microscopics of T-folds.
\subsection{Example 1 : A uniform flux}\label{linear}
If we spread the NS5-branes uniformly over the three-torus
parameterized by $x^{7,8,9}$, then we generate a uniform magnetic NSNS
three-form flux on the three-torus. If we smear the charge equivalent of $N$
NS5-branes on the three-torus residing at $x^6=0$, then the harmonic
function is (up to a constant, see e.g. \cite{Ellwood:2006my}):
\begin{equation}
h=\frac{1}{2}N\left(x^6+|x^6|\right)+c \ ,
\end{equation}
with first and second derivatives given by
\begin{equation}
\partial_{6}h=N\Theta(x^6) \ , \ \ \partial^2_{6}h=N\delta(x^6) \ .
\end{equation}
By spreading a six-dimensional object over three transverse directions
 in ten-dimensional space-time we have created a domain wall at
 $x^6=0$. On either side of the domain wall, the topology of the
 ten-dimensional space is given by six-dimensional Minkowski space
 times a three-torus $T^3$. We have taken the three-torus to have fixed volume
 $c^{3/2}$ on the left (for $x^6<0$), and to the right the volume of
 the $T^3$ evolves along the positive $x^6$-axis:
 ${\rm{V}}_{T^3}=\left(Nx^6+c\right)^{3/2}$.

\begin{figure}[htp]
\centering
\includegraphics[scale=0.9]{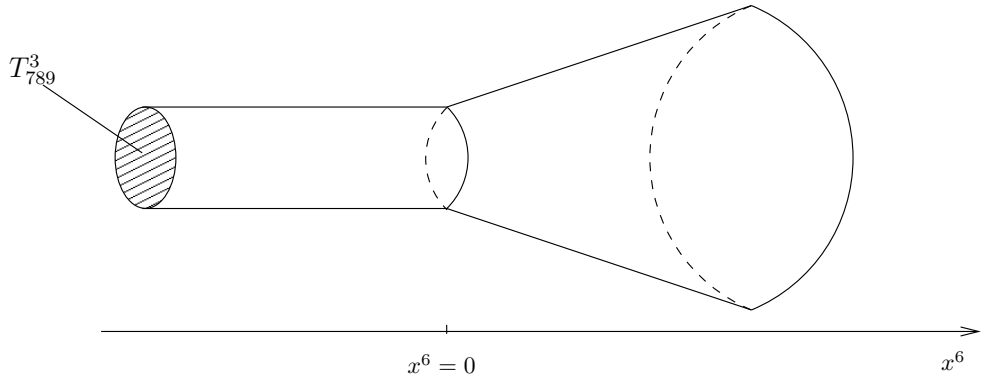}
\caption{At the location of NS5-branes spread on a three-torus, the evolution of the 
volume of the three-torus in the transverse direction changes.}
\end{figure}

The scalar curvature (see equation (\ref{scalarcurvature})) is:
\begin{equation}
R=\frac{12N^2\Theta(x^6)}{ (Nx^6+N|x^6|+2c)^3}-\frac{3N}{c^2}\delta(x^6)  \ .
\end{equation}
\subsubsection*{Remarks}
The space-time is not asymptotically flat. It behaves much like a pure
D8-brane background in type IIA string theory. In that case, it is
known that one can obtain a space T-dual to an asymptotically flat
space by including two $O8^-$ planes at the end of space-time, to
create a configuration (type I') that is T-dual to type I string
theory. To stabilize our three-torus at both infinities (on the line
transverse to the domain wall) and to obtain an asymptotically flat space-time, 
we need to include orientifold
objects with negative tension and NSNS magnetic charge. We have no
microscopic description of these objects yet although they have been
argued to exist (by using the fact that certain string theory
backgrounds should consistently describe the physics of supersymmetric
gauge theories) (see e.g. \cite{Hanany:1999sj}). 
We can think
of our background as being valid locally, near a given domain wall.

Secondly, we must check the domain of validity of our supergravity
solution, as well as the domain in space-time in which the string
coupling is small, such that our perturbative solution (in both the
string coupling and the string length over the curvature radius) is
valid.  It is clear from the supergravity solution that with an appropriate
choice of the constant $c$, and when restricting to a particular domain in $x^6$,
the supergravity solution will be valid.

Both these points illustrate the fact that it is important to
demonstrate that a given T-fold survives when backreaction is taken
into account, namely, as a full solution to weakly curved perturbative
string theory (or beyond).  From the above arguments, we decide that
the standard three-form flux case (without adding RR-fluxes) is a
borderline case in the sense that it is hard to embed it in asymptotically flat 
string theory.

\subsection*{The uniform twisted torus}
We choose the NSNS two-form $B_{(2)}$ to be:
\begin{eqnarray}
B_{(2)} &=& N x^7\Theta{\left(x^6\right)} dx^8 \wedge dx^9,
\end{eqnarray}
and perform the T-duality transformation in the $x^8$-direction to get
the uniform twisted torus \cite{Hull:1998vy}:
\begin{eqnarray}
ds^2&=&ds^2_{\mathbb{R}^{5,1}}+h\Big(\left(dx^6\right)^2+\left(dx^7\right)^2+\frac{1}{h^2}\Big(dx^8-N\Theta(x^6)x^7dx^9\Big)^2+\left(dx^9\right)^2\Big) \ ,\nonumber \\
e^{2\phi}&=&e^{2\phi_0}\ , \nonumber\\
B_{(2)}&=&0 \ .
\end{eqnarray}
After T-duality in a direction transverse to the NS5-branes, the NS5-brane charge
 disappears from the background.
The space is flat except at the point $x^6=0$ where we have a curvature
singularity as can be checked by computing:
\begin{equation}
R=-\frac{h''}{h^2}=-\frac{N}{c^2}\delta(x^6)=-\frac{N\delta(x^6)}{\det \ g} \ ,
\end{equation}
where $g$ is 10-dimensional metric. The  microscopic description available
for the singularity is that it is T-dual to the NS5-branes we started out
with. That is sufficient to interpret the backreacted twisted torus as giving rise
to a type of curvature singularity that is resolved
by string theory. Let's describe it in an alternative fashion.

\subsubsection*{The monodromy domain wall}
The presence of the domain wall at the point $x^6=0$ can also be
measured in another way. At the domain wall, there is a change in
monodromy of the twisted torus  \cite{Hull:1998vy}. In other words, we have a {\em
monodromy domain wall}. Measuring the difference of the monodromy on
either side of the domain wall is a geometric equivalent of the
measurement of the difference in the flux through the three-torus on
either side of the NS5-brane in the original background. Let's
demonstrate this in detail.

It is sufficient to consider the transverse space spanned by the
coordinates $x^{6,7,8,9}$. For $x^6<0$ there is no monodromy in the
three-torus fiber as we go around the $x^7$ cycle. On the other side
of the domain wall, for $x^6>0$, we find a 
monodromy as we go around the $x^7$ cycle given by
\begin{equation}
\left(
\begin{array}{ccc}
1&0&0\\
0&1&-N\\
0&0&1
\end{array}
\right).
\end{equation}
 The monodromy matrix has a non-trivial action only on the two-torus
 $x^{8,9}$ and as such it is an element of $SL(2,{\mathbb Z})$. It is a parabolic
 element, which is already in the canonical upper diagonal form (which
 is unique), and we can therefore uniquely associate the number $N$ to
 our twisted torus. The charge of the monodromy domain wall is
 $N$. More generically, if we allow twisted tori with parabolic
 monodromies on either side of the domain wall, then the charge of the
 monodromy domain wall is given by the difference in the numbers $N_L$
 and $N_R$ associated to the parabolic monodromies to the left and the
 right of the domain wall. Thus we see that the backreacted twisted
 torus codes the charge of the microscopic object in a geometric
 fashion.

It could be interesting to consider twisted tori with other types of monodromies,
and to analyze the properties of the monodromy domain walls between them.

\subsection*{The uniform T-fold}
After performing a second T-duality along the $x^9$-direction we
obtain the T-fold:
\begin{eqnarray}
ds^2&=&ds^2_{\mathbb{R}^{5,1}}+h\left(\left(dx^6\right)^2
+\left(dx^7\right)^2+\frac{\left(dx^8\right)^2+\left(dx^9\right)^2}{h^2
+\left(Nx^7\Theta\left(x^6\right)\right)^2}\right) \ ,\nonumber\\
e^{2\phi}&=&\frac{he^{2\phi_0}}{h^2+\left(Nx^7\Theta\left(x^6\right)\right)^2}\ , \nonumber\\
B_{(2)} &=&-\frac{Nx^7}{h^2+\left(Nx^7\Theta\left(x^6\right)\right)^2} dx^8
\wedge dx^9.
\end{eqnarray}
The covering space of the three-torus is no longer invariant under
translations in the $x^7$-direction. The curvature has also lost its
status of gauge invariant observable -- it is no longer
well-defined on the torus:
\begin{equation}
R=\frac{12N^2\Theta(x^6)}{(Nx^6+N|x^6|+2c)^3}+\frac{N}{c^2}
\left(1-4\frac{\left(Nx^7\right)^2}{c^2+\left(Nx^7\right)^2}\right)\delta
\left(x^6\right) \ .
\end{equation}
We note in particular that the curvature depends explicitly on the
periodic coordinate $x^7$. Despite the fact that the flux was
uniformly spread on the three-torus (in the directions $x^{7,8,9}$, we
have a non-trivial dependence on the $x^7$ coordinate only. Let's see in a little
 more detail how this came about.

\subsection{A note on  Wilson surfaces}
In the original geometric background, we can measure the gauge invariant observables:
\begin{eqnarray}
W_{kl} & =& e^{  2 \pi i \int_{T_{kl}} B},
\end{eqnarray}
where $k,l$ range over the coordinates of the three-torus $x^{7,8,9}$.
These are well-defined for a gerbe (see e.g. \cite{Hitchin:1999fh}), since the two-form 
$B_{(2)}$ is shifted
by 
 the curvature of a line bundle under a gauge transformation. 

A first application of the fact that these Wilson surfaces are gauge invariant
is that  two-forms $B$ of the form:
\begin{eqnarray}
B^1_{(2)} &=&  N x^7 d x^8 \wedge dx^9 \ ,
\nonumber \\
B^2_{(2)} &=&  N x^8 d x^9 \wedge dx^7
\end{eqnarray}
are gauge equivalent on $\mathbb{R}^3$ (where there are no non-trivial compact
two-cycles), but they are inequivalent on the three-torus. In particular, we
can measure the Wilson surfaces along two out of the three directions
$x^{7,8,9}$ and we find that these take different values for the two choices
of $B_{(2)}$ field, thus proving the
inequivalence of the backgrounds. 
In  particular, only the first choice of two-form is consistent
with the demand that all gauge invariant observables be invariant under
translations in the $x^{8,9}$ directions.
This observation explains why
the doubly T-dual T-fold depends on the $x^7$ direction, and not on
the true isometric directions $x^{8,9}$. 

A further use of these Wilson surface observables is as follows.
We can add
the following constant two-forms to the $B$-field:
\begin{eqnarray}
B^{extra}_{(2)} &=& b_{8} \, d x^9 \wedge dx^7 + b_{9} \, dx^7 \wedge dx^8,
\end{eqnarray} 
since they do not carry extra energy.
Since we can measure the constants $b_{8,9}$ (modulo an integer), these
backgrounds with non-trivial surface holonomies are inequivalent to
the background we studied before. After T-duality, they generate new
twisted tori and T-fold backgrounds.  It is straightforward to apply
the Buscher rules to obtain explicit formulas for the metric, dilaton
and NSNS two-form in these backgrounds.

To make that point more concrete, we believe it is 
sufficient to study the standard T-fold case without backreaction:
\begin{eqnarray}
ds^2&=&(dx^7)^2+(dx^8)^2+(dx^9)^2 \ , \nonumber\\
B_{(2)}&=&Nx^7 dx^8 \wedge dx^9+B^{extra}_{(2)} \ .
\end{eqnarray}
After a  T-duality transformation along $x^8$ the metric and two-form become
\begin{eqnarray}
ds^2&=&\left(1+b_9^2\right)(dx^7)^2+(dx^8-Nx^7 dx^9)(b_9dx^7+dx^8-Nx^7dx^9)+(dx^9)^2 \ ,\nonumber\\
B_{(2)}&=&b_8 dx^9\wedge dx^7,
\end{eqnarray}
and after an additional T-duality along $x^9$
\begin{eqnarray}
ds^2&=&(dx^7)^2+\frac{1}{1+(Nx^7)^2}\left(\left(dx^9-b_8 dx^7\right)^2+\left(dx^8+b_9 dx^7\right)^2\right) \ ,\nonumber\\
B_{(2)}&=&\frac{Nx^7}{1+(Nx^7)^2}\left( -dx^8 \wedge dx^9+b_9 dx^9\wedge dx^7 - b_8 dx^7 \wedge dx^8\right) \ .
\end{eqnarray}
One can also effortlessly produce inequivalent backreacted T-folds
following this strategy of introducing surface holonomies.

\subsubsection*{Classical gauge invariants}
We have generated backreacted geometric, twisted tori and T-fold
backgrounds.  Since applying the Buscher rules transforms all the
(gauge variant) objects determining these backgrounds (like the metric,
and NSNS two-form) it is natural to ask how the gauge invariant
objects are mapped into one another under such a
transformation. 

One route towards defining classical gauge invariant objects as measured
in a given background solution is the
following.  We consider a gauge invariant combination $O[g,B,\dots]$
of the fields in the original geometric background (e.g. the Ricci
scalar 
or the three-form flux $H_{(3)}$ at a given point
in space-time). We then apply T-duality to
the object in the sense that we rewrite the gauge invariant as a
functional of the T-dual fields $\tilde{g}, \tilde{B}$
etcetera. Clearly, the dual will be a complicated expression in
the T-dual variables, but by T-duality, it will remain a gauge
invariant object. The disadvantages of this formulation of gauge
invariant objects in T-folds are on the one hand that it leads to
unwieldy expressions and, more importantly, that it is only available
when we have a geometric dual. We can address these points by looking
on the one hand for expressions that are invariant in form under
T-duality transformations. On the other hand and more importantly,
 we would like to have an
intrinsic definition of gauge invariants in T-folds that is independent of
the existence of a geometric dual.
We are then looking for gauge invariants
that are not only invariant under coordinate transformations, but also under
the T-duality transformations that occur when we change patch in a T-fold.
Such objects should be invariants not only of geometric gauge 
transformations, but also of the T-duality group. 

In the following, we want to give an example of how one can
formulate a solution to both problems in practice.
 Consider the moduli
 fields $\rho$ and $\tau$ of the two-torus $T^2_{89}$ on which we
 performed T-duality transformations in our first example.
The T-dualities we consider act by
 $O(2,2,\mathbb{Z})$ transformations on the pair of moduli.  These
 include $SL(2,\mathbb{Z}) \times SL(2,\mathbb{Z})$ transformations,
 as well as the exchange of the two moduli.
Thus, if we consider an unordered
 pair of modular invariant $j$-functions of the two moduli:
\begin{equation}
(j(\rho),j(\tau)) \ ,
\end{equation}
then we have classical gauge invariants that are independent of the T-duality frame
in which we study the backgrounds. That addresses the first issue.

Note however, that it also gives a technical solution to the second
issue. If in a given T-fold we change patch, we act by an
$O(2,2,\mathbb{Z})$ transformation on the moduli fields, and again the
set of numbers is invariant, now under a change of coordinate
patch. Thus, the $O(2,2,\mathbb{Z})$ invariant that we constructed can
be used to define gauge invariants in T-folds, intrinsically.
The generalization of this example to bigger T-duality or U-duality groups 
should be clear.

After this digression on classical gauge invariants, let's turn to a second example.

\subsection{Example 2: Localized flux}
We can generalize the constant flux example, while improving our
control on the gravitational backreaction. We have already explained
(via the measurement of Wilson surfaces) that we only have true
isometries in two directions of the three-torus. We can make further
use of this freedom to localize the NS5-brane source in the $x^7$
direction.  The harmonic function is then of the form
\begin{equation}\label{h}
h(x^6,x^7)=\frac{N}{8\pi}
\log\left( \sinh^2\left(\pi x^6\right)+\sin^2\left(\pi x^7\right)\right)
\end{equation}
and fulfills
\begin{equation}\label{laplace}
\Delta h(x^6,x^7)=N\delta(x^6)\delta_{\mathbb{Z}}(x^7) \ ,
\end{equation}
where $\delta_{\mathbb{Z}}$ denotes the periodic delta-function. The harmonic
function codes the backreaction to $N$ NS5-branes which sit at the point
$x^6=0, \ x^7=0$ (and $x^7$ is compact). In the example of the linear harmonic
function $h$ the singularity was of co-dimension
one, producing a domain wall. The singularity is now of co-dimension two, so
it is a vortex. More precisely, it corresponds to six-dimensional objects
spread on a two-torus, and localized on $\mathbb{R} \times S^1$.

We can measure the presence of the NS5-branes by measuring their magnetic
charge
under the NSNS three-form $H_{(3)}$ by taking an integral over the
$H_{(3)}$-field around the point $x^6=x^7=0$:
\begin{eqnarray}\label{charge}
\int_{C\times T_{89}}H_{(3)}=\oint_C\left(\partial_{6}hdx^7-\partial_{7}hdx^6\right)=\int_{-\infty}^{+\infty}dx^6\int_{0}^{1}dx^7 \Delta h =N \ ,
\end{eqnarray}
where $C$ is the curve circling the vortex on the $x^{6,7}$ cylinder.
The equations of motion and the curvature can be read off from the
formulas in section \ref{general}. The $H_{(3)}$-field varies over the 
three-torus:
\begin{eqnarray}
H_{(3)}&=&\frac{N}{4}\frac{\sin(2 \pi x^7)}{\cos(2\pi x^7)-\cosh(2\pi x^6)}
dx^6 \wedge dx^8 \wedge dx^9 \ ,\nonumber\\
&-&\frac{N}{4}\frac{\sinh(2\pi x^6)}{\cos(2\pi x^7)-\cosh(2\pi x^6)} dx^7
\wedge dx^8 \wedge dx^9 \ .
\end{eqnarray}
The T-duality transformation along $x^8$ gives us
(via the formulas of section \ref{general}) a background 
with twisted torus topology:
\begin{equation}
ds^2=ds^2_{{\mathbb R}^{5,1}}+h\left(\left(dx^6\right)^2+\left(dx^7\right)^2+\frac{1}{h^2}\left(dx^8-bdx^9\right)^2+\left(dx^9\right)^2\right) \ 
\end{equation}
with the function $b$ given by
\begin{equation}\label{bb}
b=\int_0^{x^7}  \partial_{x^6}h dx'^7=-\int_0^{x^6}\partial_{x^7} h dx'^6=\frac{N}{4\pi}\arctan \left(\frac{\tan(\pi x^7)       }{\tanh(\pi x^6)}\right) \ .
\end{equation}
Let us determine which branch of the arctangent
function we should take. We can determine this by noting that at
infinity, the localized NS5-brane on the cylinder cannot be
distinguished from the circularly spread density of NS5-branes that we
had before. Thus, at large value of $x^6$, the solution should agree
with the uniform solution.

\begin{figure}[htp]
\centering
\includegraphics[scale=0.7]{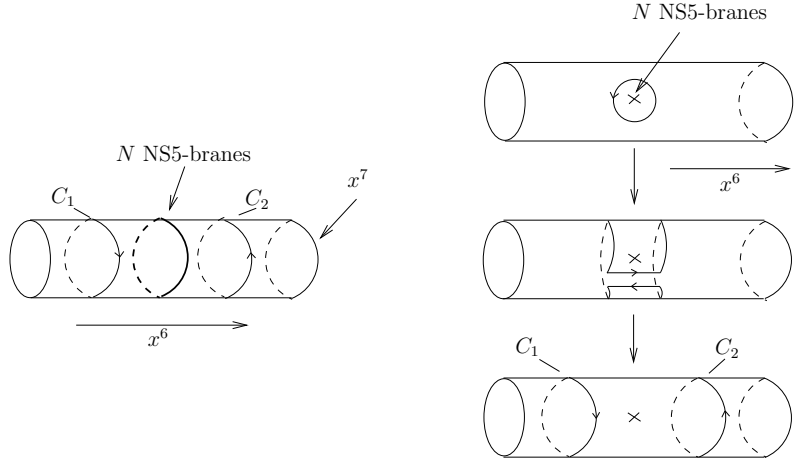}
\caption{Far from the NS5-branes, the uniform and localized distributions on the three-torus
match.}\label{charge_mod}
\end{figure}

Therefore the asymptotics of
$b$ must be given by the following choice of branches:
\begin{equation}
\left.b\right|_{x^6=\pm \infty}=\pm \frac{ N x^7}{4} \ .
\end{equation}
\subsubsection*{Monodromy vortex}
As we discussed in detail previously, far from the source we will see it as a
monodromy domain wall.
However, we know that we should now be able to localize the source more
precisely. We are therefore lead to define an observable that gives a more
refined measurement of the geometric singularity (than the monodromy of the
twisted
torus around the $x^7$ cycle).

We know that the $H_{(3)}$-flux in the original background is a
derivative of the real part of the K\"ahler modulus. By T-duality
transformation the K\"ahler modulus is mapped to a complex structure
modulus. That suggests that we should be able to measure the presence
of a {\em monodromy vortex } 
in the derivative of the complex structure modulus. The
monodromy vortex characterizes a new kind of 
twisted torus geometry.  Let's see
how this works in practice. We denote the real part of the complex
structure modulus $\tau_1={\text{Re}}(\tau)$. Then
we can compute the vortex monodromy as follows:
\begin{equation}
\oint_{C_{67}}d\tau_1=\oint\left(\partial_{6}\tau_1 dx^6+\partial_{7}\tau_1
  dx^7\right)
=\oint\left(\partial_{7}h dx^6-\partial_{6}h dx^7\right)=-N \ ,
\end{equation}
where we used that
\begin{equation}
\tau_1=-b=-\int_{0}^{x^7} dx'^7h 
\partial_{dx^6}=\int_{0}^{x^6} dx'^6h\partial_{dx^7} \ .
\end{equation}

\subsubsection*{Remark}
The monodromy vortex was discussed in a slightly different guise
in \cite{Hellerman:2002ax}, and it is familiar from other contexts. For instance, it is akin to
 the monodromy in the dilaton-axion field that is
generated by the D7-brane in type IIB string theory. For that matter, it is a
phenomenon quite familiar from the backreaction due to any co-dimension two
object governed by a Laplace equation. More specifically, here we find a
monodromy in the complex structure which is different from a monodromy in the
dilaton-axion. However, in F-theory we can code the monodromy of the D7-brane
in a monodromy of the complex structure of an auxiliary two-torus. The
difference is that here, the monodromy is in the complex structure
modulus of a two-torus that is part of the physical ten-dimensional space-time.
From our discussion it becomes manifest that the discussion of \cite{Hellerman:2002ax} of the monodromy
vortex pertains to a full supergravity solution, corresponding to NS5-branes spread on a two-torus.

\subsubsection*{Doubly T-dual T-fold}
We can also study how the NSNS flux is coded in the doubly T-dual T-fold. Since the K\"ahler modulus $\tilde{\tilde\rho}$ of the doubly T-dual T-fold  satisfies
\begin{equation}
\tilde{\tilde\rho}=-\frac{1}{\rho} \ ,
\end{equation}
where $\rho$ is the K\"ahler modulus of the original (geometric)
background, the magnetic charge of the NS5-brane which we computed in
equation (\ref{charge}) can be written as
\begin{equation}
N=-\oint {\rm Re} \, \frac{1}{\tilde{\tilde\rho}} \ .
\end{equation}
The charge we computed in this way is not the canonical NSNS charge associated
to the three-form flux $H_{(3)}$. 
This procedure provides an example of how an
observable $O$ can be literally translated into a dual background, as
we discussed previously.

\subsection*{Remark}
We note that the NS5-brane spread on a two-torus only has better backreaction properties than the uniform flux
example. We only logarithmically differ from an asymptotically flat background instead of linearly. As such, one can 
attempt to compactify the space transverse to the NS5-branes by combining a sufficient number of individual sources
to restore the total curvature of a two-sphere. That was done  in \cite{Hellerman:2002ax} by globally
gluing approximations to the local solutions presented here.

\section{Non-geometric regions in configuration space}
Until now we have discussed examples of T-folds which have a geometric
dual. If we take the point of view that in the path integral of string
field theory (namely, second quantized string theory) we should divide
out by the full gauge group which includes not only diffeomorphisms
but also T-duality (or U-duality) transformations, then the points of
configuration space that we considered up to now are automatically
included in an integral over geometric configurations \footnote{For
the sake of simplicity we ignore the exchange of for instance type IIA with IIB
string theory under T-duality. The reader can imagine
that we discuss bosonic string theory.}.

In this section we would like to study whether we can find points in
the configuration space of string field theory that have no geometric
equivalent in their gauge orbit. There are some constructions of
such points in the literature, which includes half K3 manifolds glued in a particular
way \cite{Hellerman:2002ax}, as well as 
asymmetric orbifold points  \cite{Narain:1986qm}. 
We will discuss a new such point in configuration space with
distinctive features in the next section.
In any case, it is good to make those points more manifest,
since it is in these new
regions of the configuration space of string theory that the construction
 of T-folds (or U-folds) becomes most useful. 

We want to  show that other regions of configuration space
exist that are not integrated over when considering only geometric
backgrounds. In a first step, we will not worry about whether the point we construct is
a solution to the equations of motion, since our main goal is to
show that we must integrate in an (off-shell) path integral over more
than only geometric backgrounds. 

We first concentrate on the following subproblem: can we construct a
point in configuration space that has no geometric U-dual.  
It is intuitively clear that such points exist.
When we glue patches via duality transformations, and then act on the
U-fold with local gauge transformations patch by patch, and global duality transformations, 
we will not generically be able to trivialize all gluings.


To make this more concrete, let's concentrate on T-folds, and T-duality transformations.
Our construction will be as follows. We consider a two-torus fibration
over a circle. As we go around the circle (with coordinate $x^7$), the two-torus can pick up a
monodromy $M$ in the T-duality group. When we appropriately choose the
monodromy, we demonstrate that it cannot be T-dualized to a geometric
monodromy. We can summarize the problem at hand in the following diagram:
\begin{equation}
\begin{CD}
\Big(\rho,\tau\Big) @>x^7\rightarrow x^7+1>>M \cdot 
\Big(\rho,\tau\Big)\\
@VV {D}V  @VV {D} V    \\
D \cdot \Big(\rho,\tau\Big) @>x^7\rightarrow x^7+1
>>
\left( D \cdot M \cdot D^{-1} \right) \cdot D \cdot 
\Big(\rho, \tau \Big).\\
\end{CD}
\end{equation}
We need to show that we can choose a monodromy $M$ which is
non-geometric such that for any T-duality $D$ the new monodromy $D
\cdot M \cdot D^{-1}$ is also non-geometric.

Firstly, we consider a monodromy $M$ to be geometric if it factorizes
on the K\"ahler and complex structure modulus, and if it is moreover
of the type $T^n$ for the K\"ahler modulus (where $T$ is the operator
that shifts the K\"ahler modulus by one). In other words, the only
geometric monodromies for the K\"ahler modulus are shifts by an
integer $n$. For the complex structure any $SL(2,{\mathbb Z})$
transformation is an ordinary (geometric) global diffeomorphism.  We
note therefore that a K\"ahler structure monodromy
is  of
parabolic type when geometric. 
When we conjugate the geometric
monodromy, we will always remain with a parabolic monodromy. We also
recall that a T-duality transformation can act to exchange
K\"ahler and complex structure modulus. To avoid geometrization of the
model via the transport of the non-geometric K\"ahler monodromy to a
geometric complex structure monodromy, we must also demand that the
complex structure monodromy is not of parabolic type. (In this discussion
we have excluded the special case of a constant modulus which lies at the
fixed point of a non-trivial monodromy.)

It is therefore sufficient to choose a model with  non-parabolic
monodromies for both the K\"ahler and the complex structure modulus in
order to have a model which cannot be T-dualized to a geometric
background. Such a model is a point in a new non-geometric region of
configuration space.

Many explicit examples can be constructed (see e.g. \cite{Flournoy:2004vn}). We give one example.
Consider a model with monodromies 
\begin{eqnarray}
\rho(x^7+1)&=&S\cdot \rho(x^7)=-\frac{1}{\rho(x^7)} \ , \nonumber\\
\tau(x^7+1)&=&S\cdot \tau(x^7)=-\frac{1}{\tau(x^7)} \ .
\end{eqnarray}
A possible realization for the $\rho$-modulus would be of the
following kind. Let
\begin{equation}
P_1=\{x^7| 0<x^7<1\} \ \ {\rm and} \ \ P_2=\{x^7| \frac{1}{2}<x^7<\frac{3}{2}\} \nonumber
\end{equation}
be an open covering of the base circle $S^1$ and let
\begin{equation}
A=(0,\frac{1}{2}) \ \ {\rm and} \ \ 	B=(\frac{1}{2},1) \nonumber
\end{equation}
be the intersection of the two patches $U_1\cap U_2$. 
The local trivialisation $\phi_1$ and $\phi_2$ on the patches
$P_{1,2}$ are given by
\begin{equation}
\phi_1^{-1}(u)=(x^7,t) \ \ {\rm and} \ \ \phi_2^{-1}(u)=(x^7,t) \nonumber
\end{equation}
for $u$ a coordinate on the patch and
 $x^7\in A$ and $t\in T^2$. The transition function
 $t_{12}$ on the part $A$ of the intersection of patches
 is the identity map. On the other part $B$ of the intersection
the transition function is
\begin{equation}
t_{21} \ : \ \phi_1^{-1}(u)=(x^7,t) , \ \ \phi_2^{-1}=(x^7, S\cdot t) \ , \nonumber
\end{equation}
where $S$ is a generator of the T-duality group and maps coordinates
of a torus with volume $\rm{Im}(\rho)$ to coordinates of the torus
with volume $\frac{1}{\rm{Im}(\rho)}$.

\begin{figure}[htp]
\centering
\includegraphics[scale=0.9]{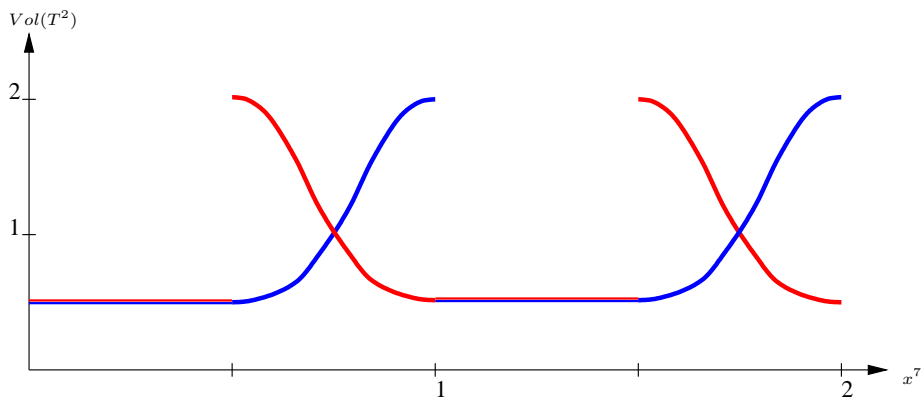}
\caption{We draw an example of an evolving modulus with elliptic monodromy.}
\end{figure}

Therefore, it is not too hard to find regions in configuration space
that are truly non-geometric.  However, in a second step, we must
take into account the vacuum selection done by the equations governing
string backgrounds. In particular, when we choose an elliptic
monodromy, as we did above and we assume that the moduli only depend
on the compactification direction $x^7$, then the moduli will tend to
relax to constant values, and in particular, for an elliptic
monodromy, the moduli relax to the fixed point of the elliptic
monodromy matrix. At these fixed points, then, the elliptic monodromy
becomes equivalent to a trivial monodromy (since the modulus is
constant). The backgrounds corresponding to these moduli have an enhanced
discrete symmetry \cite{Dabholkar:2002sy}\cite{Flournoy:2004vn}. 
The discrete symmetry can then be used to (asymmetrically) orbifold the background to make
it non-geometric \cite{Narain:1986qm}.

Note also that once a modulus stabilizes
at its fixed point value, it can be interpreted as a modulus with monodromy, 
or a modulus with trivial monodromy. In other words, those are points in moduli
space were a T-fold topology change could occur. 
 The difference between the two interpretations 
lies in the spectrum of allowed fluctuations. It would be interesting to see whether one
can argue for such a T-fold monodromy/topology change transition.

Finally, when we consider constant moduli
with hyperbolic and parabolic Scherk-Schwarz ansatz, then we find that these do
not provide us with fixed points -- the potentials (without gradient terms)
exhibit runaway behavior \cite{Dabholkar:2002sy}.

We can now learn an important lesson from the study of the doubly
T-dual to the NS5-brane solution. It provides us with a background
with parabolic monodromy, with a modulus that varies over space.
Moreover, the solution is stable (and preserves sixteen supercharges).
Therefore we are lead to search for new non-geometric backgrounds that 
allow for a modulus that varies over space, in order to find new non-geometric backgrounds 
that lie outside the reach of attractive fixed points.

\section{A new space-dependent solution}
The solutions we studied in detail in the first sections, have a 
duality twist from the parabolic conjugacy
class of $SL(2,{\mathbb Z})$. From the
analysis of \cite{Dabholkar:2002sy}, we know that when we reduce the supergravity equations
of motion to seven dimensions after reducing on $T^2$ and
additionally on a circle with parabolic or hyperbolic duality twists, then there exists
no stable constant minimum in the resulting potential.

Since we have a concrete solution, namely the doubly T-dual of NS5-brane solutions,
 which is stable (since it is supersymmetric) and which has 
a parabolic duality twist, it is interesting to analyze  how we can  generalize 
the analysis of \cite{Dabholkar:2002sy} in order to include that type
of solution. In doing so, we may learn how to construct interesting solutions of a
different type altogether. At the very least, we will find an alternative to the relaxation of the moduli
to constant fixed point values.

\subsection{The equations of motion in seven dimensions}
In this subsection we briefly remind the reader of how dimensional reduction with duality
twists proceeds (see e.g. \cite{Meessen:1998qm,Hull:2002wg}). We concentrate on the part of the
eight-dimensional Lagrangian that contains the complex and K\"ahler moduli describing
the geometry of the two-torus on which we compactify.
Additionally, we recall that Scherk and Schwarz considered compactifications with fields which
depend on the compactified directions \cite{Scherk:1979zr}. We will reduce the eight-dimensional
action further (along the $x^7$-direction) 
to seven dimensions using such a Scherk-Schwarz reduction. The dependency of the fields
on the $x^7$ direction will be such that it drops out of the eight-dimensional Lagrangian,
rendering a further dimensional reduction straightforward. The consistency of the reduction scheme
was understood in \cite{Scherk:1979zr}.
%
Concretely, the ten-dimensional
fields do not depend on $x^8$ and $x^9$ directions along the two-torus $T^2$, and after reducing
we have (amongst others) two additional scalar fields, namely the K\"ahler modulus
$\hat{\rho}$ and the complex structure modulus $\hat{\tau}$ of the $T^2$-fiber. In
the reduced eight-dimensional Lagrangian they transform under
$SL(2,{\mathbb Z})_{\hat{\rho}}\times SL(2,{\mathbb Z})_{\hat{\tau}}$. Next, one
Scherk-Schwarz reduces the eight-dimensional fields along the angular $x^7$ direction.

The relevant terms in the eight-dimensional action for the moduli are the $SL(2, \mathbb R)$ invariant
$SL(2,\mathbb R)/U(1)$ coset actions:
\begin{equation}
S_{mod}^{(8)}=\int d^8 x\sqrt{g} e^{-2\phi^{(8)}}\left(-\frac{\partial_m \hat \rho\partial^m\hat{\bar\rho}}{\hat \rho_2^2}-\frac{\partial_m \hat \tau\partial^m\hat{\bar\tau}}{\hat \tau_2^2}\right) \ .
\end{equation}
The dilatons in eight and ten dimensions are related by the formula
\begin{equation}
\phi^{(8)}=\phi^{(10)}-\frac{1}{4}\log \left(\det g_{T^2_{89}}\right) \ .
\end{equation}
We can rewrite the action in the form
\begin{equation}\label{star}
S^{(8)}_{mod}=\frac{1}{2}\int d^8 x\sqrt g e^{-2\phi^{(8)}} Tr\left(\partial_m \hat H^{-1}\partial^m\hat H\right) \ ,
\end{equation}
where we take the moduli field $\hat{H}$ to have the form: 
\begin{equation}\hat H
=\frac{1}{\hat \rho_2}\left(\begin{array}{cc}1&\hat \rho_1\\ \hat\rho_2&|\hat \rho|^2\end{array}\right)  \oplus \frac{1}{\hat \tau_2}\left(\begin{array}{cc}1&\hat \tau_1\\ \hat\tau_2&|\hat\tau|^2\end{array}\right) \ .
\end{equation}
We consider a Scherk-Schwarz ansatz for the moduli that
guarantees that the $x^7$ dependency will drop out in the Lagrangian:
\begin{equation}
\hat H(x^7)={\cal M}^T(x^7)H{\cal M}(x^7)=e^{M^Tx^7}He^{M x^7}.
\end{equation}
The unhatted field $H$ no longer depends on the angular coordinates $x^7$. The exponential
factors give a monodromy to the moduli of the $T^2_{89}$ fiber.
Inserting this ansatz into the action (\ref{star}) we obtain the seven-dimensional reduced action:
\begin{eqnarray}\label{actionH}
&&\int d^7x \sqrt {g^{(7)}}e^{-2 \phi^{(7)}} \, 
Tr \, \left(\frac{1}{2}\partial_m H^{-1}\partial^{m}H-g^{77}\left(M^2+M^THMH^{-1}\right)\right)
\end{eqnarray}
with $m=0,\ldots,6$ and $\phi^{(7)}=\phi^{(8)}-\frac{1}{4}\log
g_{77}$. In the following we further reduce our ansatz and assume that
there is no non-trivial monodromy in the complex structure modulus. We consider
only the T-duality transformation and monodromies that act upon the K\"ahler modulus only.
The action is then classically invariant under $SL(2, \mathbb{R} )$
duality transformations. These act on the matrices 
%
$H=\frac{1}{\rho_2}\left(\begin{array}{cc}1&\rho_1\\
    \rho_1&|\rho|^2\end{array}\right)$ and $M$ as follows:
\begin{equation}
H\longrightarrow A^THA \ \ \ {\rm and} \ \ \ M\longrightarrow A^{-1}HA \ ,
\label{transfoprop}
\end{equation}
where $A$ is an $SL(2, \mathbb R)$ matrix.
We now recall the action for monodromy matrices $m$ in various conjugacy classes
of $SL(2,\mathbb R)$. For the monodromy matrix from the parabolic conjugacy class
\begin{equation*}
M_p=\left(\begin{array}{cc}0&m\\0&0\end{array}\right)
\end{equation*}
we obtain the seven-dimensional action:
\begin{equation}
S_{mod}^{(p)}=-\int d^7x\sqrt{g^{(7)}}e^{-2\phi^{(7)}}\left(\frac{m^2 g^{77}+\partial_m\rho\partial^m\bar \rho}{\rho_2^2}\right) \ .
\end{equation}
For the mass matrix from the elliptic conjugacy class
\begin{equation*}
M_e=\left(\begin{array}{cc}0&m\\-m&0\end{array}\right)
\end{equation*}
we obtain
\begin{equation}
S_{mod}^{(e)}=-\int d^7x\sqrt{g^{(7)}}e^{-2\phi^{(7)}}\left(\frac{m^2 g^{77}|1+\rho^2|^2+\partial_m\rho\partial^m\bar\rho}{\rho_2^2}\right) \ ,
\end{equation}
and for the mass matrix from the hyperbolic conjugacy class
\begin{equation}
M_h=\left(\begin{array}{cc}m&0\\0&-m\end{array}\right)
\end{equation}
we obtain
\begin{equation}
S_{mod}^{(h)}=-\int d^7x
\sqrt{g^{(7)}}e^{-2\phi^{(7)}}
\left(\frac{4m^2 g^{77}|\rho|^2+\partial_m\rho\partial^m\bar \rho}{\rho_2^2}\right) \ .
\end{equation}
In the following we further assume that the K\"ahler modulus is constant
along the $x^0,\ldots , x^5$-directions. In contrast to  \cite{Dabholkar:2002sy}, we allow 
for a dependence of the moduli on the  $x^6$-direction. As a result, when analyzing solutions to the equations
of motion we not only take into account the potential, but also the gradient terms.
The equations of motions which we derive from the above
actions are
\begin{eqnarray} \label{eom}
&&(\rho-\bar\rho)\left(\partial_m\partial^m\bar\rho+\left(\sqrt{g^{(7)}}e^{-2\phi^{(7)}}\right)^{-1}\partial_m  \left(\sqrt{g^{(7)}}e^{-2\phi^{(7)}}\right)\partial^m\bar \rho\right) +2\partial_m\bar \rho\partial^m\bar \rho
\nonumber \\
& & 
+2m^2 g^{77}=0 \ ,\nonumber\\
&&(\rho-\bar\rho)\left(\partial_m\partial^m\bar\rho+\left(\sqrt{g^{(7)}}e^{-2\phi^{(7)}}\right)^{-1}\partial_m  \left(\sqrt{g^{(7)}}e^{-2\phi^{(7)}}\right)\partial^m\bar \rho\right) +2\partial_m\bar \rho\partial^m\bar \rho\nonumber \\
& & +2m^2 g^{77}\left(1+\bar \rho^2\right)\left(1+|\rho|^2\right)=0 \ ,\nonumber\\
&&(\rho-\bar\rho)\left(\partial_m\partial^m\bar\rho+\left(\sqrt{g^{(7)}}e^{-2\phi^{(7)}}\right)^{-1}\partial_m  \left(\sqrt{g^{(7)}}e^{-2\phi^{(7)}}\right)\partial^m\bar \rho\right) +2\partial_m\bar \rho\partial^m\bar \rho
\nonumber \\
& & 
+4m^2 g^{77}\bar \rho(\rho+\bar\rho)=0\nonumber\\
\end{eqnarray}
for the parabolic, elliptic and hyperbolic conjugacy classes respectively.

\subsection{A space-dependent modulus with parabolic monodromy}
We have tuned our ansatz such that the doubly T-dual solution of
section \ref{linear} falls inside the class. We can thus explicitly
check on that example the equations of motion, and verify that indeed
one finds a spatial dependence of the modulus that gives rise to the
desired monodromy.  The $SL(2, \mathbb R)$ invariant gradient terms
cancel out the (otherwise runaway) potential terms to provide new
solutions to the equations of motion.  Explicitly, the K\"ahler
modulus of the solution is given by
\begin{equation}
\hat\rho_1=-\frac{Nx^7}{(Nx^6+c)^2+(Nx^7)^2} \ , \qquad \hat\rho_2=\frac{Nx^6+c}{(Nx^6+c)^2+(Nx^7)^2} \ .
\end{equation}
The monodromy which we read of from its behavior along the angular $x^7$-direction is
\begin{equation}
\tilde M=\left(\begin{array}{cc}0&0\\-N&0\end{array}\right)=A^{-1}\left(\begin{array}{cc}0&N\\0&0\end{array}\right)A \ \ \ \  {\rm with}   \ \ \ \ A=\left(\begin{array}{cc}0&-1\\1&0\end{array}\right) \ .
\end{equation}
This matches with the  Scherk-Schwarz ansatz:
\begin{equation}
\frac{1}{\hat\rho_2}\left(\begin{array}{cc}1&\hat\rho_1\\ \hat \rho_1&|\hat\rho|^2\end{array}\right)=A^T e^{\left(\begin{array}{cc}0&0\\N&0\end{array}\right)x^7}  \left( \begin{array}{cc}\frac{1}{Nx^6+c}&0\\0&N x^6+c\end{array}   \right)  e^{\left(\begin{array}{cc}0&N\\0&0\end{array}\right)x^7} A \ .
\end{equation}
Let's understand why this provides a solution to the equations of motion. In the background at hand, we have that $\phi^{(8)}$ is constant, and that the metric in $\mathbb{R}^{5,1}$ is trivial.
Moreover $g_{66}=g_{77}$ such that in the $x^6$ dependent gradient term, the non-trivial
mixing with the metric drops out completely.
If we then take the real part of the K\"ahler modulus $\rho$ to be zero and
 keep the imaginary part $\rho_2$ to have a generic $x^6$ dependence, than
 the equation of motion (\ref{eom}) simply becomes
 \begin{equation}
 \rho_2\partial_6^2\rho_2-\left(\partial_6\rho_2\right)^2+m^2=0 \ .
 \end{equation}
 The linear function $\rho_2=m x^6+c$ (with $m=N$) indeed solves the
 equation of motion of the seven-dimensional action. We have learned
 in this example that the gradient terms can  compensate for
 runaway behavior in the potential for a parabolic monodromy.
Decoupling of the equations of motion for the K\"ahler modulus follows from a specific
metric ansatz.

\subsection{On the  existence or not  of a geometric T-dual}
Both a parabolic and a hyperbolic monodromy matrix do not have a fixed point.
They necessarily give rise to non-constant moduli fields. The parabolic
monodromy gave rise to a solution that is T-dual to a geometry with
flux. One can wonder whether one can find solutions with hyperbolic
monodromy, especially in the light of the fact that we argued
previously that those cannot be T-dual to geometric backgrounds (when we restrict
the action of the T-duality group to be $SL(2, \mathbb{R})$ only).
Before we attempt to find such a solution, we revisit the analysis of
the existence of a geometric dual in the language of the
lower-dimensional field theory.

The duality transformation behavior of the moduli field
$\hat H$ can be used to confirm our discussion about
existence/non-existence of the geometric T-dual. We take the working
definition that a
globally non-geometric background implies that the volume of the
$T^2$-fibration is a non-periodic function of the base-coordinate
$x^7$. For a given globally non-geometric background a test of the
existence of the geometric T-dual works as follows.

For a given solution one writes down the matrix 
\begin{equation}
\hat H_{\rho}=\frac{1}{\hat\rho_2(x^7)} 
\left(
\begin{array}{cc}
1               &  \hat\rho_1(x^7) \\ 
\hat\rho_1(x^7) & |\hat\rho(x^7)|^2 
\end{array}
\right)
\end{equation} 
where $\hat \rho_1$ gives the value of the $B_{8,9}$ component and $\hat
\rho_2$ the volume in the given $T$-duality frame.
A conjugation of the monodromy matrix by a general $SL(2,{\mathbb
  R})$-matrix will generate an equivalent background but  with a
different expression for $\hat\rho_2$ (see equation (\ref{transfoprop})). If it possible to find such a $SL(2,{\mathbb
  R})$-matrix that the new $\hat\rho_2$ is $x^7$-independent then a
geometric T-dual does exist. One can analyze these conditions generically
for the various types of $SL(2, \mathbb{R})$ conjugacy classes, and we find the following
results:
%
%
\begin{itemize}
\item
There is no $SL(2, \mathbb R)$ transformation that transforms away a hyperbolic monodromy 
along the angular $x^7$ direction.
 \item For the elliptic conjugacy class, the dependence on the angular coordinate
$x^7$ is non-trivial unless the modulus is at the fixed point of the monodromy.
\item For a parabolic monodromy, there is a duality frame in which the modulus is
independent of the angular direction $x^7$. 
\end{itemize}

\subsection{A space-dependent modulus  with hyperbolic twist }
We now turn to finding a solution to the equations of motion (\ref{eom}) in the case where we have a hyperbolic duality twist.
Equipped with the equation (\ref{eom}) we can guess a ten-dimensional
solution with duality twist coming from the hyperbolic conjugacy
class.
When we have 
vanishing $B$-field (and therefore a purely imaginary K\"ahler modulus $\rho$) the equation
 (\ref{eom}) is solved by a constant $\rho=iC$. That gives rise to the two-torus
geometry coded in  
\begin{equation}
\hat H= \frac{1}{\hat \rho_2}\left( \begin{array}{cc}1&\hat\rho_1\\ \hat \rho_1& |\hat \rho|^2 \end{array}               \right)=
{\cal M}^T \left(\begin{array}{cc}\frac{1}{\rho_2}&0\\ 0&\rho_2\end{array}\right) {\cal M} \ \  {\rm with} \  \ {\cal M}={e^{\left(\begin{array}{cc}m&0\\0&-m\end{array}\right)x^7}} 
\end{equation}
or in other words
\begin{equation}
ds^2_{89}=C e^{-2 m x^7}\left(\left(dx^8\right)^2+\left(dx^9\right)^2\right) \ , \ \ B_{(2)}=0 \ .
\end{equation} 
Our ansatz for the other metric components is based on the fact that we only
expect an $x^6$ dependence of the other fields and metric components, and we moreover
are inspired by the relations between these fields in the parabolic solution.
Thus, we make the ansatz that $\phi^{(8)}$ only depends on $x^6$, and that 
$g_{66}=g_{77}$ only depends on the $x^6$ coordinate as well. We moreover
take $g_{67}=0=B_{(2)}$. We summarize these proposals in the expression:
\begin{eqnarray}
ds^2&=&ds^2_{{\mathbb R}_{1,5}}+h(x^6)\left(\left(dx^6\right)^2+\left(dx^7\right)^2\right) +ds^2_{89} \ ,\nonumber\\
\phi^{(10)}&=&\frac{1}{2}\log (C e^{-2 m x^7}) +\phi^{(8)}(x^6) \ .
\end{eqnarray}
We then plug this ansatz directly into the ten-dimensional equations of motion, and find with
some effort that they are solved by
\begin{eqnarray}
ds^2&=&ds^2_{{\mathbb R}^{1,5}}+h\Big(\left( dx^6\right)^2+\left(dx^7\right)^2\Big)+C e^{-2m x^7}\Big(\left( dx^8\right)^2+\left(dx^9\right)^2\Big) \ ,\nonumber\\
h&=&\frac{B}{2 x^6+A}e^{-\frac{1}{4}m^2(2x^6+A)^2}\ , \nonumber\\
\phi&=& \phi_0 +\frac{1}{2}\log\left(\frac{Ce^{-2 m x^7}}{2 x^6+A}\right) \ ,\nonumber\\
B_{(2)}&=&0 \ .
\end{eqnarray}
with $A, B, C,\phi_0$ constants.

Let us analyze the solution in slightly more detail. We note that 
for $m=0$ we obtain the metric and dilaton
\begin{eqnarray}
ds^2&=&ds^2_{{\mathbb R}^{1,5}}+\frac{B}{2 x^6+A}\Big(\left( dx^6\right)^2+\left(dx^7\right)^2\Big)+C \Big(\left( dx^8\right)^2+\left(dx^9\right)^2\Big) \ ,\nonumber\\
\phi&=& \phi_0 +\frac{1}{2}\log\left(\frac{C}{2 x^6+A}\right) \ ,
\end{eqnarray}
which in the new coordinate system $z=\sqrt{B(A+2x^6)}$ reduces to:
\begin{eqnarray}
ds^2&=&ds^2_{{\mathbb R}^{1,5}}+ dz^2+\frac{B^2}{z^2}\left(dx^7\right)^2+C \Big(\left( dx^8\right)^2+\left(dx^9\right)^2\Big)\nonumber\\
\phi&=& \phi_0 +\frac{1}{2}\log\left(\frac{BC}{z^2}\right) \ , \qquad e^{\phi}=\frac{e^{\phi_0}\sqrt{BC}}{z} \ .
\end{eqnarray}
A T-duality along the $x^7$-direction 
\begin{eqnarray}
ds^2&=&ds^2_{{\mathbb R}^{1,5}}+ dz^2+\frac{z^2}{B^2}\left(dx^7\right)^2+C \Big(\left( dx^8\right)^2+\left(dx^9\right)^2\Big) \nonumber\\
 e^{\phi}&=&e^{\phi_0} \sqrt\frac{{C}}{{B}} 
\end{eqnarray}
shows that the metric without monodromy is T-dual to an (almost everywhere) flat background.
If we wish to avoid a conical singularity at $z=0$, we must tune the parameter $B$ appropriately.

For a non-zero hyperbolic monodromy, our solution is non-trivial. It
cannot be brought into a geometric frame with an $SL(2, \mathbb{R})$
duality transformation, and the curvatures are non-trivial. It has a
certain domain of validity in which both the curvatures and the string
coupling constant are small. The singularity that the original
solution exhibits is of a type T-dual to a flat or conical space. 
It would be good to check the properties of these solutions further,
and in particular to study their stability through a fluctuation
analysis that properly takes into account the T-fold boundary
conditions.

Note also that we have exhibited the solution in a form which is appropriate for hyperbolic
monodromies in the full $SL(2,{\mathbb R})$ group. It is straightforward 
to bring it into a form suitable for all $SL(2,{\mathbb Z})$-valued twists with $|Tr({\cal M})|>2$. 
These are of two types of hyperbolic $SL(2, \mathbb{Z} )$ conjugacy classes, namely the generic ones
with representatives:
\begin{eqnarray}
{\cal M} &=&
\left(
\begin{array}{cc}
n & 1
\\
-1 & 0
\end{array}
\right)
\end{eqnarray}
where $n$ is an integer with absolute value larger than three, and sporadic conjugacy classes that 
one can enumerate.
Let us give us an example of how to construct a solution with such a monodromy in practice.
Consider for example a solution with sporadic 
monodromy ${\cal M}(8)=\left(\begin{array}{cc}1&2\\3&7\end{array}\right)$. We will obtain
a classically equivalent solution
 if we set $m=\log(4-\sqrt{15})$ in our solution. Additionally, we can generate infinitely
 many solutions with this conjugacy class by  
$SL(2,{\mathbb R})$-conjugation of the monodromy matrix, and in particular there are many frames
in which the monodromy is indeed $SL(2, \mathbb Z)$ valued. 
Note that we can also use duality rotations to generate solutions with hyperbolic monodromy and non-trivial
NSNS three-form $H_{(3)}$.
In summary, we determined a new solution to the equations of motion
which has non-trivial varying K\"ahler modulus that exhibits a hyperbolic
monodromy.

To motivate the subsequent subsection, we note that we could turn a
background of this form in type IIA/B string theory into a background
with hyperbolic monodromy in the complex structure modulus of IIB/A
string theory, thus rendering the monodromy geometric. We use a
T-duality transformation outside the $SL(2,\mathbb{Z})_{\hat{\rho}}$ duality group
to achieve this.  It should be clear from our previous discussions
that the way to avoid such geometrization in a mirror geometry, we
need to introduce a non-trivial (say hyperbolic) monodromy for the
complex structure as well. Can we find a supergravity solution with a hyperbolic monodromy in both
the K\"ahler and complex structure ?

\subsection{Let's twist again}
Indeed, we found a supergravity solution with
a non-trivial monodromy in both the K\"ahler and the complex
structure modulus. The underlying reason for the simplicity of the
generalization is that the monodromies of both K\"ahler and complex structure modulus
enter the dynamics of the other metric components and the dilaton in a similar fashion.
The solution for the metric, dilaton and NSNS two-form $B_{(2)}$ is as follows:
\begin{eqnarray}
ds^2&=&ds^2_{{\mathbb R}^{1,5}}+h\Big(\left( dx^6\right)^2+\left(dx^7\right)^2\Big)+C e^{-2m_1 x^7}\left( dx^8\right)^2+C e^{-2m_2 x^7}\left(dx^9\right)^2\ ,\nonumber\\
h&=&\frac{B}{2 x^6+A}e^{-\frac{(m_1^2+m_2^2)(2x^6+A)^2}{8}}\nonumber\\
\phi&=& \phi_0 +\frac{1}{2}\log\left(\frac{Ce^{- (m_1+m_2) x^7}}{2 x^6+A}\right)\ ,\nonumber\\
B_{(2)}&=&0 \ .
\end{eqnarray}
From these one learns immediately that the 
K\"ahler and complex structure modulus are given by:
\begin{eqnarray}
\hat\rho=B_{89}+i\sqrt{g_{T^2_{89}}}=iC e^{-(m_1+m_2) x^7} \ , \ \ \ \ \hat\tau=\frac{g_{89}}{g_{88}}+i\frac{\sqrt{g_{T^2_{89}}}}{g_{88}}= i e^{(m_1-m_2) x^7}\ .
\end{eqnarray}
Rewriting the moduli fields using the
$\hat H$-matrix allows us to identify the type of monodromy for the above solution.
\begin{eqnarray}
\hat H_{\rho}&=&\left(\begin{array}{cc}C^{-1}e^{(m_1+m_2) x^7}&0\\0&Ce^{-(m_1+m_2) x^7}\end{array}\right)={\cal M}_\rho^T\left(\begin{array}{cc}\frac{1}{C}&0\\0&C\end{array}\right){\cal M}_\rho\nonumber \ ,\\
\hat H_{\tau}&=&\left(\begin{array}{cc}e^{(m_2-m_1) x^7}&0\\0&e^{(m_1-m_2) x^7}\end{array}\right)={\cal M}_\tau^T\left(\begin{array}{cc}1&0\\0&1\end{array}\right){\cal M}_\tau
\end{eqnarray}
with
\begin{equation}
{\cal M}_\rho= e^{\left(\begin{array}{cc}\frac{m_1+m_2}{2}&0\\0&-\frac{m_1+m_2}{2}\end{array}\right)x^7}   \ , \ \ \ \ {\cal M}_\tau=e^{\left(\begin{array}{cc}\frac{m_2-m_1}{2}&0\\0&\frac{m_1-m_2}{2}\end{array}\right)x^7} \ .
\end{equation}
For $m_1\neq m_2$ and $m_1\neq -m_2$ we have hyperbolic monodromies in
both sectors. The solution is genuinely non-geometric under all
$O(2,2,{\mathbb Z})$ duality transformations. We can tune the two
hyperbolic parameters and use the $O(2,2,{\mathbb R})$ duality group
to construct the solutions for which the hyperbolic monodromies take
values in $O(2,2,{\mathbb Z})$, as we illustrated in the previous subsection.

\section{Conclusion}
We have given the gravitational backreaction of T-folds T-dual to
purely NSNS background. It transpires that twisted tori and T-folds
correspond to new types of gravitational singularities which are
resolved via T-duality and known resolutions. We extended the analysis
to cases with Wilson surfaces and flux on a three-torus
localized in one direction. The concept of monodromy domain walls and
vortices is useful to describe the microscopic origin of twisted
tori. We showed for the importance of including the full backreaction of 
proposed T-folds in order to judge whether they can be defined in an asymptotically
flat string theory.

Moreover, we argued that interesting non-trivial non-geometric backgrounds exist in which we allow the
moduli to vary over non-compact space. In fact, the doubly T-dual to a NS5-brane is an example of such 
a background which is geometrizable. We found a supergravity solution with hyperbolic monodromies which 
is not equivalent to a geometric one.
It will be interesting to further analyze the properties of the solution, and in particular to analyze its stability through a fluctuation analysis that properly takes into account the T-fold boundary conditions.

Thus we showed with an explicit example that one can find regions in the
configuration space of second quantized string theory that are
non-geometric.  It would be good to study these regions further and to estimate to what
degree their contributions to a second quantized string theory path
integral are important. We expect that they may be of 
importance for instance in cosmological big crunch big bang scenarios and in 
string phenomenology.

\section*{Acknowledgments}
We would like to thank all members of the \'Ecole Normale Sup\'erieure -
 Jussieu study group and in particular Costas Bachas, Raphael Benichou
 and Atish Dabholkar for discussions. Our work was supported in part by the
EU under the contract MRTN-CT-2004-005104 and by the ANR (CNRS-USAR) contract
05-BLAN-0079-01.

\end{document}